# Crystal fields and Kondo effect: specific heat for Cerium compounds

**H.-U. Desgranges[1]**

[1] scientifically unaffiliated, Albert-Kusel-Str. 25, Celle, Germany

E-mail: H-Ulrich.Desgranges@nexgo.de

The thermodynamic Bethe ansatz equations for the N = 6 Coqblin-Schrieffer model with crystal fields have been solved numerically. The realistic case of three Kramers doublets with arbitrary splittings has been studied for the first time. The specific heat has been calculated for representative combinations of the ionic energy splittings providing ample material for comparison with experimental results for Cerium impurities and compounds.

**1. Introduction**

The single ion Kondo model and its generalization to a N-fold degenerate ionic configuration, the SU(N) Coqblin-Schrieffer[1] model, has been used successfully to describe the thermodynamic properties of dense Kondo systems[2,3,4]. Apparent non-Fermi-liquid like behavior has been explained (within a N = 4 approximation) as stemming from the interplay of Kondo effect and crystal field splittings[5,6].

The change of the effective spin-degeneracy N due to the interplay between Kondo and crystal field effects has been studied experimentally by investigating various Cerium based pseudo-ternary intermetallic substitution series[7]. Specific heat data have been fitted by combining exact N = 2 Bethe ansatz or resonant level model results with crystal field Schottky terms[8]

Within the exact solution[9,10] of the Coqblin-Schrieffer model by Bethe ansatz the specific heat was calculated already for three special cases of crystal field configurations: (a) the case that the N=6 multiplet is split into a $\Gamma_6$ doublet and a $\Gamma_8$ quartet by small cubic crystal fields[11], (b) the case that a low lying quartet is further split into two doublets while the highest doublet can be neglected, resulting in an effective N=4 model[12], and (c) the case that the N=6 multiplet is split into three equidistant Kramers doublets[13] (cf. Fig. 1).

In the two latter cases it was possible to treat the full range of crystal field splittings showing in the specific heat the separation for large crystal fields into an effective spin-1/2 (N=2) Kondo peak at low temperatures and a Schottky-type peak at high temperatures. In all three cases a shoulder like structure develops for low to intermediate crystal fields. In the case of a low lying $\Gamma_8$ quartet the shoulder appears at the high temperature side of the peak.

In the present work the above cases have been generalized to the case of three Kramers doublets with arbitrary splittings. Also case (a) has been reexamined numerically and extended to higher crystal field strengths.

These new results allow a quantitative comparison with experimental data that makes it possible to identify deviations from single-ion behavior. They also show how the interplay between crystal fields and Kondo effect reduces the effective spin-degeneracy N.

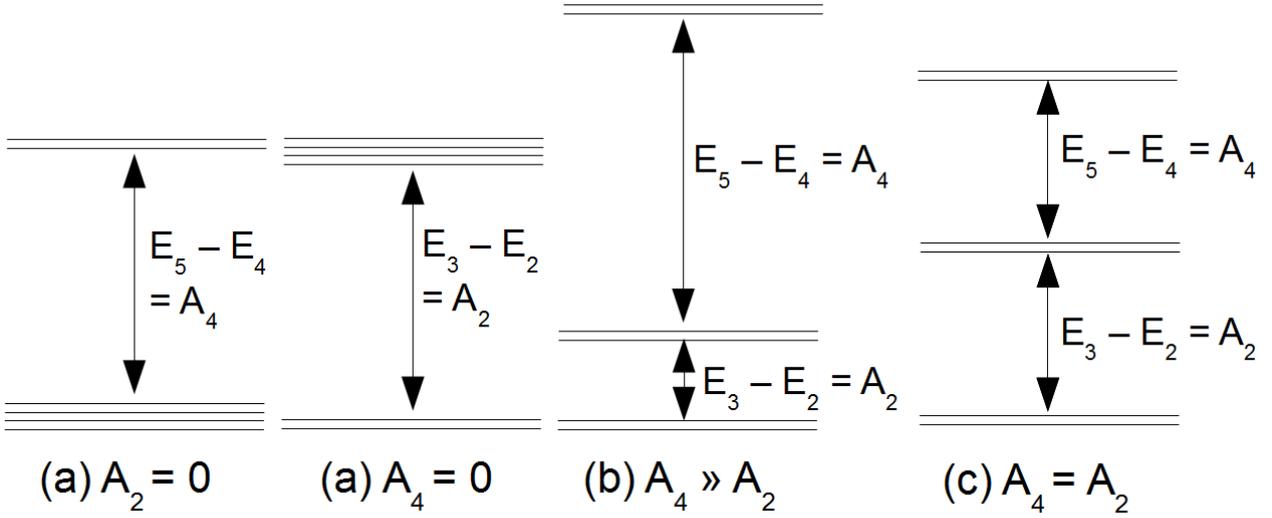

Fig. 1. The energy levels for the ionic ground state of $Ce^{3+}$ ions in cubic crystal fields: case (a) and in non-cubic crystal fields considered in cases (b) $A_4 \gg A_2 \equiv A$ and (c) $A_4 = A_2 \equiv A$.

## 2. Model and thermodynamic equations

The Coqblin-Schrieffer Hamiltonian can be written in terms of the N ionic crystal field states $|r\rangle$ with energy levels $E_r$ and the usual notation for conduction electron operators $C^\dagger_{k,r}$ where the exchange interaction is given by a permutation operator:

$$H = \sum_{k,r} k\, C^\dagger_{k,r} C_{k,r} + J \sum_{k,r;k',r'} |r\rangle\langle r'| C^\dagger_{k',r'} C_{k,r} + \sum_r E_r |r\rangle\langle r| \qquad (1)$$

For integrability of the model a linear dispersion of the conduction electron energy is assumed as well as a small exchange coupling $J$ independent of $r$. The Bethe ansatz solution requires also an ad hoc cut-off D that enters the Kondo temperature (non-universally defined here as) $T_K \sim D \exp(-1/N|J|)$.

The thermodynamic properties of the model are calculated from certain pseudo-energy functions $\varepsilon_n^{(r)}(\lambda)$, $n = 1, 2,..., \infty$, $1 \leq r \leq N-1$ that are determined by the Bethe ansatz equations[9].

In the scaling limit $J \to 0$, $D \to \infty$, $T_K$ kept fixed these read (with $\varepsilon_0^{(r)} = -\infty$):

$$-\ln\{1 + \exp[-\varepsilon_n^{(r)}(\lambda)/T]\} = -\sin(r\pi/N)\exp[\lambda]\delta_{n,1}$$
$$+ \sum_{q=1}^{N-1} S_q^r * (\ln\{1+\exp[\varepsilon_{n+1}^{(q)}(\lambda)/T]\} + \ln\{1+\exp[\varepsilon_{n-1}^{(q)}(\lambda)/T]\} - s^{-1} * \ln\{1+\exp[\varepsilon_n^{(q)}(\lambda)/T]\}), \qquad (2)$$

where $s * f(\lambda)$ denotes the convolution $s * f(\lambda) = \int_{-\infty}^{\infty} s(\lambda - \lambda') f(\lambda') d\lambda'$, and the kernels $S_q^r$ are given by their Fourier transforms:

$$S_q^r(\omega) = \frac{\sinh(\min(q,r)\pi\omega/N)\,\sinh((N-\max(q,r))\pi\omega/N)}{\sinh(\pi\omega)\,\sinh(\pi\omega/N)} \quad \text{and} \quad s^{-1}(\omega) = 2\cosh(\pi\omega/2).$$

The free energy at temperature T is given by the following expression:

$$F = -T \sum_{r=1}^{N-1} \int_{-\infty}^{\infty} \frac{\sin(r\pi/N) \ln\{1+\exp[\varepsilon_1^{(r)}(\lambda)/T]\} \, d\lambda}{\{\cosh[\lambda - \ln(T_K/T)] - \cos(r\pi/N)\} 2\pi} \quad (3)$$

The thermodynamic properties in the absence of external fields depend only on the ratio $T/T_K$. $T_K \equiv T_K(N)$ can be related by the universal Wilson number[14] to the conventionally defined Kondo temperature. It is connected with the linear specific heat coefficient $\gamma = C/T$ for $T \to 0$ through $T_K(N) = (N-1)\pi/(3\gamma)$.

The numerical solution of the type of equations (2) has become standard practice since their first occurrence[15] with the spin-1/2 (N=2) Kondo model, provided that the limiting values $\varepsilon_n^{(r)}(\lambda)$ for $\lambda \to \pm\infty$ are known. For $n \geq 2$ the equations can be written in the following form[10]:

$$\begin{aligned}\varepsilon_n^{(r)}/T &= s*\ln\{1+\exp[\varepsilon_{n+1}^{(r)}(\lambda)/T]\} + s*\ln\{1+\exp[\varepsilon_{n-1}^{(r)}(\lambda)/T]\} \\ &\quad - s*\ln\{1+\exp[-\varepsilon_n^{(r+1)}(\lambda)/T]\} - s*\ln\{1+\exp[-\varepsilon_n^{(r-1)}(\lambda)/T]\}.\end{aligned} \quad (4)$$

In the limit $\lambda \to -\infty$ the integral equations reduce to the following algebraic recurrence relations (with the notation: $g_r^{(r)} \equiv \ln\{1+\exp[\varepsilon_n^{(r)}(-\infty)/T]\}$, $b_n^{(r)} \equiv -\ln\{1-\exp[-g_n^{(r)}]\}$):

$$\begin{aligned}&g_n^{(r)} - \tfrac{1}{2}\{g_{n+1}^{(r)} + g_{n-1}^{(r)}\} = b_n^{(r)} - \tfrac{1}{2}\{b_n^{(r+1)} + b_n^{(r-1)}\}, \; g_0^{(r)} = 0, \; b_n^{(0)} = b_n^{(N)} = 0 \\ &\lim_{n \to \infty} g_{n+1}^{(r)} - g_n^{(r)} = A_r/T.\end{aligned} \quad (5)$$

where the generalized fields $A_r$ ($A_r \geq 0$) are related to the energy levels $E_r$ of the ionic ground state in the particular crystal field configuration: $A_r = E_{r+1} - E_r$, $1 \leq r \leq N-1$, cf. Fig. 1.

An analytic solution to equations (5) is known only in the magnetic field case ($A_r = g\mu_B H$ for all $r$). In the above cases (b) and (c) a numerical solution was accomplished by an interval halving method[13]. This was facilitated by the fact that in both cases only one field parameter $A_2 \equiv A$ has to be taken into account (cf. Fig. 1, $A_1 = A_3 = A_5 = 0$), and that due to the symmetry $r \to N-r$ of equations (5) the problem reduces to a two-dimensional one in case (b) and a three-dimensional in case (c). In the general case of unequal Kramers splittings $A_4 \neq A_2$ the problem is five-dimensional and this method appears not to be feasible.

In order to overcome this problem a new strong field (or low temperature) expansion has been devised for the case (c) that has then been generalized to the new case (d) of three Kramers doublets with unequal splittings ($A_4 \neq 0$, $A_2 \neq 0$) and to case (a) ($A_2 = 0$ or $A_4 = 0$). The physical backing for this expansion comes from the observation that for large values of the crystal field splitting the system can essentially be described as an effective spin-1/2 (N=2) Kondo system (or an effective spin-3/2 (N=4) system for the case (a) with $A_2 = 0$).

We look exemplarily at the cases (c) and (d) describing three Kramers doublets. Writing $A_4 = x A$, $A \equiv A_2$ ($x > 0$), for $A/T \to \infty$ we observe the decoupling of equations (5) due to $b_n^{(2)} \to 0$, $b_n^{(4)} \to 0$ for $A/T \to \infty$. The solution to (5) is then

$$\begin{aligned}g_n^{(t)}(A/T \to \infty) &= 2\ln(n+1), \; t = 1,3,5 \\ g_n^{(s)}(A/T \to \infty) &= n A_s/T - \tfrac{1}{2}\{g_n^{(s+1)}(A/T \to \infty) + g_n^{(s-1)}(A/T \to \infty)\}, \; s = 2,4\end{aligned} \quad (6)$$

We now introduce the deviations from the strong field limit as

$$g_n^r = g_n^r(A/T \to \infty) + u_n^{(r)}, \; 1 \leq r \leq 5 \quad (7)$$

and rewrite equations (5) in terms of $u_n^{(r)}$:

$$u_n^{(t)} - \tfrac{1}{2}\{u_{n+1}^{(t)} + u_{n-1}^{(t)}\} + \ln\{1 - k_n[\exp(-u_n^{(t)}) - 1]\} =$$
$$\tfrac{1}{2}\ln\{1 - (n+1)^2 \exp(-nA_{t+1}/T - u_n^{(t+1)})\} + \tfrac{1}{2}\ln\{1 - (n+1)^2 \exp(-nA_{t-1}/T - u_n^{(t-1)})\}, \; t = 1,3,5$$
$$u_n^{(s)} - \tfrac{1}{2}\{u_{n+1}^{(s)} + u_{n-1}^{(s)}\} + \ln\{1 - (n+1)^2 \exp(-nA_s/T - u_n^{(s)})\} = \quad (8)$$
$$\tfrac{1}{2}\ln\{1 - k_n[\exp(-u_n^{(s+1)}) - 1]\} + \tfrac{1}{2}\ln\{1 - k_n[\exp(-u_n^{(s-1)}) - 1]\}, \; s = 2,4$$
with $A_0 \equiv A_6 \equiv \infty$ and $k_n = [n(n+2)]^{-1}$.

Equations (8) allow for a systematic expansion of $u_n^{(r)}$ in powers of exp(-A/T) and are also amenable to a numerical treatment by iteration.

From this expansion we have come to the following formulae for the general case (d) confirmed by numerical evaluation:

$$g_n^{(t)} = 2\ln[n+1+\alpha^{(t)}] + O(\exp(-(n+1)\text{Min}(A_s)/T)), \; t = 1,3,5$$

$$\alpha^{(1)} = \frac{-2\exp(-A_2/T)}{1-\exp(-A_2/T)} + \frac{-2\exp(-(A_2+A_4)/T)}{1-\exp(-(A_2+A_4)/T)}$$

$$\alpha^{(3)} = \frac{-2\exp(-A_2/T)}{1-\exp(-A_2/T)} + \frac{-2\exp(-A_4/T)}{1-\exp(-A_4/T)}$$

$$\alpha^{(5)} = \frac{-2\exp(-A_4/T)}{1-\exp(-A_4/T)} + \frac{-2\exp(-(A_2+A_4)/T)}{1-\exp(-(A_2+A_4)/T)} \quad (9)$$

$$g_n^{(s)} = nA_s/T - \tfrac{1}{2}\{g_n^{(s+1)} + g_n^{(s-1)}\} + f_s(A_2/T, A_4/T) + O(\exp(-(n+1)\text{Min}(A_s)/T)), \; s = 2,4$$

$$f_2(A_2/T, A_4/T) = -4\ln[1-\exp(-A_2/T)] + 2\ln[1-\exp(-A_4/T)] - 2\ln[1-\exp(-(A_2+A_4)/T)]$$

$$f_4(A_2/T, A_4/T) = 2\ln[1-\exp(-A_2/T)] - 4\ln[1-\exp(-A_4/T)] - 2\ln[1-\exp(-(A_2+A_4)/T)]$$

We can determine $g_n^{(r)}$ numerically if we first truncate equations (8) at some large value m, solve them for $u_{n-1}^{(r)}$, then insert $g_{m+1}^{(r)}$ and $g_m^{(r)}$ according to (9) via (7) and compute all $u_n^{(r)}$ with $0 \leq n \leq m-1$ down the chain; $u_0^{(r)}$ computed this way will in general not be equal to 0. Next we set $u_0^{(r)} = 0$ and use the computed $u_n^{(r)}$ with $1 \leq n \leq m-1$ as starting values that we insert into the original form of the truncated equations (8) and calculate $u_n^{(r)}$, $1 \leq n \leq m$ up the chain. We take the obtained values of $u_n^{(r)}$ as new starting values for the next iteration and repeat the procedure until the relative difference between two iterations is of the desired accuracy.

The case (a) that the N=6 multiplet is split into a $\Gamma_6$ doublet and a $\Gamma_8$ quartet has been treated similarly. However, in this case the corrections to $g_n^{(r)}$ do not vanish exponentially but only $O(1/n^2)$.

With these results at hand the numerical solution of the thermodynamic Bethe ansatz equations has been achieved for a number of crystal field splittings. The specific heat has been calculated by numerical differentiation of the free energy. The relative accuracy is expected to be better than 1%. The symbols in the depicted specific heat curves indicate the finite temperature grid.

## 3. Results for the specific heat

First we note that for equal crystal field splittings, i.e. case (c) the results of ref. 13 have been recovered for the limiting values $g_n^{(r)}$ and by extension for the specific heat as well.

In comparison with ref. 11, where case (a) is considered, we get only qualitative agreement. Taking into account that their $|\Delta_C|/T_0$ corresponds to $A/(T_K/2)$ in our notation, we get disagreement in the height of the main peak and in the development of the second peak for values $A/T_K > 0.5$. Our results are shown in Fig. 2. Contrary to ref. 11 the second peak has already developed for $A_2/T_K(N=6) = 1.0$ (I) and $A_4/T_K(N=6) = 2.0$ (II), respectively. However, the exact numerical values in the scaling relations for the effective N=2 or N=4

Kondo temperatures given in eq. (6) and (7) of ref. 11 are reproduced numerically by the present calculations. Since the numerical procedure for solving the integral equations (2) including the determination of the limiting values (5) has not been published by the authors of ref. 11 we suppose that it is valid only for small values of the crystal field splittings.

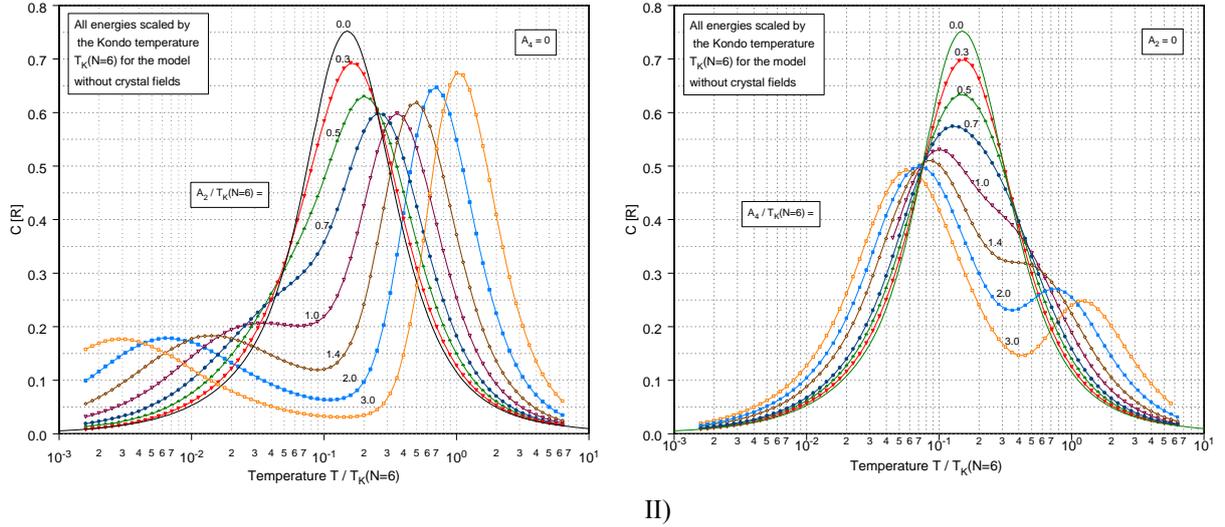

I)                                                                                              II)

Fig. 2. (Color online) Specific heat as a function of T scaled by the Kondo temperature $T_K(N=6)$ for the model without crystal fields. I) Case (a) $\Gamma_7$ doublet lower, i.e. $A_4 = 0$; II) Case (a) $\Gamma_8$ quartet lower, i.e. $A_2 = 0$.

The crystal field splittings $A_2$ and $A_4$ are scaled by the Kondo temperature in the absence of crystal fields $T_K(N=6)$. In Fig. 3 we show the specific heat C for the values of $A_2/T_K(N=6)$ equal to I) 0.3, II) 0.5, III) 0.7, and IV) 1.0. The ratio $A_4/A_2$ serves as an additional parameter. The low temperature side of the curves shows a similar behavior as in the published results[13] for $A_4 = A_2$. For increasing values of $A_2$ a shoulder develops into a peak associated with the Kondo resonance of the effective spin-1/2 system at low temperatures. With increasing values of $A_4/A_2$ a reduction in the height of the central peak is observed that is accompanied by the development of a shoulder into a third peak on the high temperature side. It is noteworthy that the second peak is centered at about the same position for varying values of $A_4/A_2$ for a given value of $A_2$ as soon as the third peak has formed.

In Fig. 4 we show the specific heat C for the values of $A_4/T_K(N=6)$ equal to I) 1.0, and II) 3.0, with the value of $A_2/T_K(N=6)$ serving as an additional parameter. Graph I) interpolates between case a) $A_2 = 0$ and case c) $A_4 = A_2$ and shows the development upon increasing $A_2$ of the two peak structure associated with the effective spin-1/2 system at low temperatures and the crystal field levels at high temperatures. Graph II) starts with the two peak structure associated with the quartet ground state and the crystal field peak at high temperatures and shows the development of the corresponding three peak structure upon increasing $A_2$ to intermediate values. Again it is noteworthy that the third peak is centered at about the same position for varying values of $A_2/T_K(N=6)$. Upon further increasing $A_2$ the second and the third peak merge.

Finally, we take a more experimentalist's look at the data. Since the model's scale $T_K(N=6)$ is unknown, we shift the specific heat curves on the logarithmic temperature axis so that the low temperature limit (linear in T) coincides with the spin-1/2 Kondo curve[15]. The corresponding curve, denoted as SU(2), is also shown.

Therefore we show in Fig. 5 the specific heat as a function of T scaled by the respective Kondo temperature $T_K(N=2)$ of the effective spin-1/2 system at low temperatures: $C = (\pi/3) T/T_K(N=2)$.

For case (a) with lower doublet and case (c) the two Kondo temperatures are related in the scaling limit

$A_2 \gg T_K(N=6)$ by the following expression[11,13]:

$$T_K^3(N=6) = C_4 A_2^2 T_K(N=2). \tag{10}$$

If we linearly extrapolate the exact numerical values for $C_4$ of cases (a) and (c) we arrive at

$$C_4 = 17.08 + 17.05\, A_4/A_2. \tag{11}$$

This relation has been verified within the accuracy of the applied procedure (5%) generally for case (d) for $A_2 \geq 0.7\, T_K(N=6)$.

If the crystal field levels are known experimentally the Kondo temperature $T_K(N=6)$ can be calculated from formulae (10) and (11).

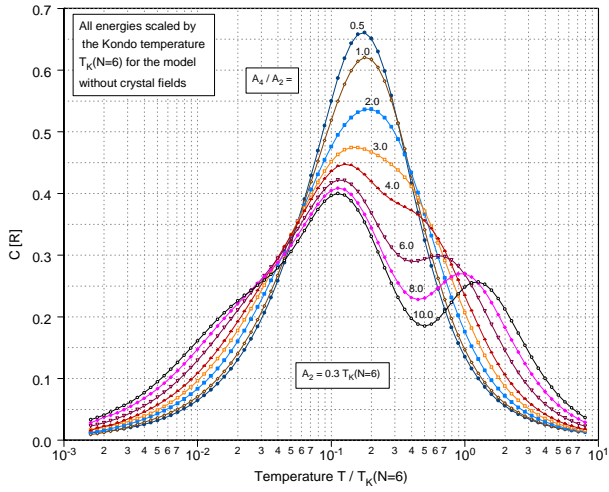

I)

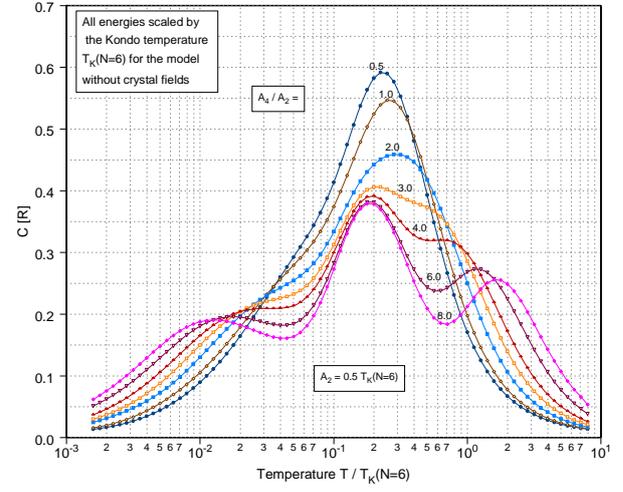

II)

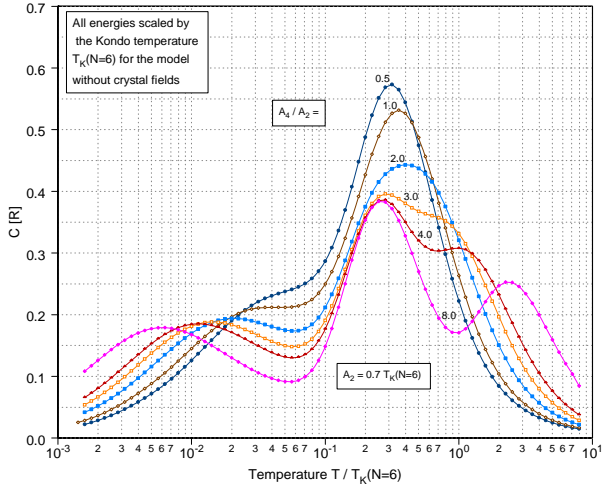

III)

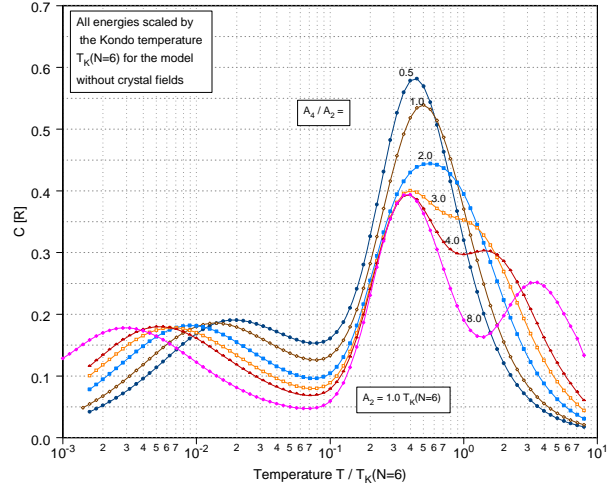

IV)

Fig. 3. (Color online) Specific heat as a function of T scaled by the Kondo temperature $T_K(N=6)$ for the model without crystal fields. $A_2/T_K(N=6)$ is kept fixed for each graph.

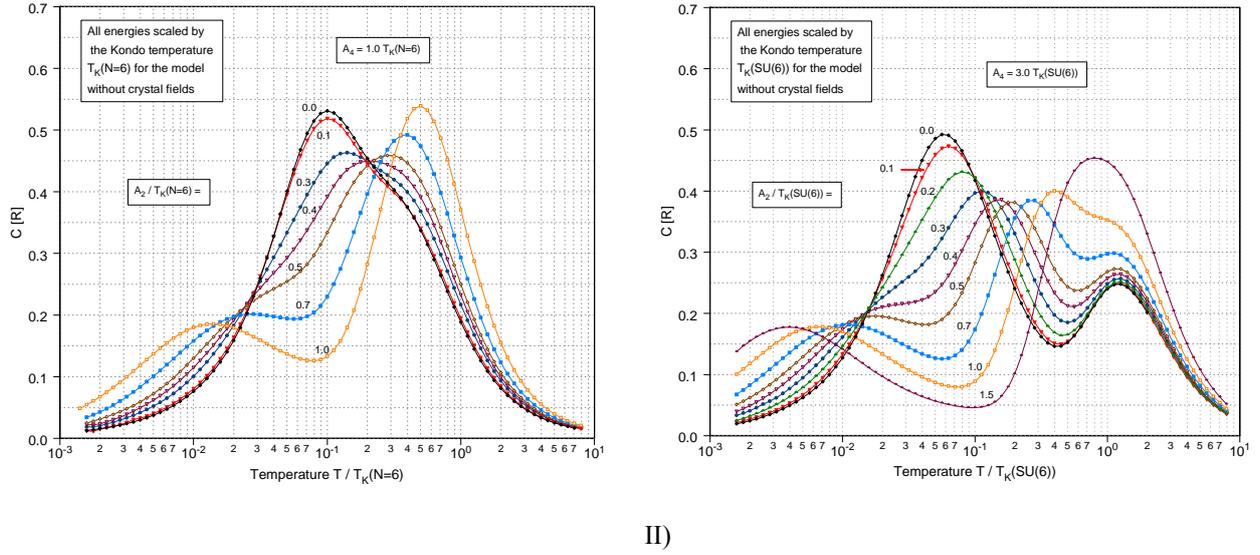

I)          II)

Fig. 4. (Color online) Specific heat as a function of T scaled by the Kondo temperature $T_K(N=6)$ for the model without crystal fields. $A_4/T_K(N=6)$ is kept fixed for each graph.

## 4. Discussion

On the basis of these results a comparison of theoretical and experimental curves can be performed without free parameters[16]. In comparison to Numerical Renormalization Group[17] (NRG) calculations the computational effort of the present calculations is rather low.

Recently a heuristic resonant level approach has been proposed to fit experimental specific heat data for Cerium compounds[18]. While the agreement with earlier exact $N = 4$ results is quite good a comparison with the present results for three doublets would be interesting.

As pointed out by these authors[18] the exact solution of the Coqblin-Schrieffer model presumes that the exchange coupling of different ionic multiplets with the conduction electrons is the same. However, for large crystal field splittings the coupling of the higher multiplets is reduced[19]. This would justify the procedure of Romero et al. where the highest multiplet is taken into account by a Schottky expression that leads to an increase in the peak height associated with the highest multiplet compared with the present work. An investigation to quantify this effect by NRG calculations would be desirable.

## 5. Conclusion and outlook

With this work all physical cases of crystal field splittings for Cerium ions (N=6) within the exact solution of the Coqblin-Schrieffer model in the absence of magnetic fields have been covered. To solve the thermodynamic equations a new strong field (or low temperature) expansion has been devised. Calculated here for the first time, a three peak structure for larger crystal field splittings has been found. Magnetic fields may be included along the same lines. Results for the zero field magnetic susceptibility for $N = 4$ will be published elsewhere[20].

## Dedication

This publication is dedicated to the memory of J. W. Rasul (1953-2011).

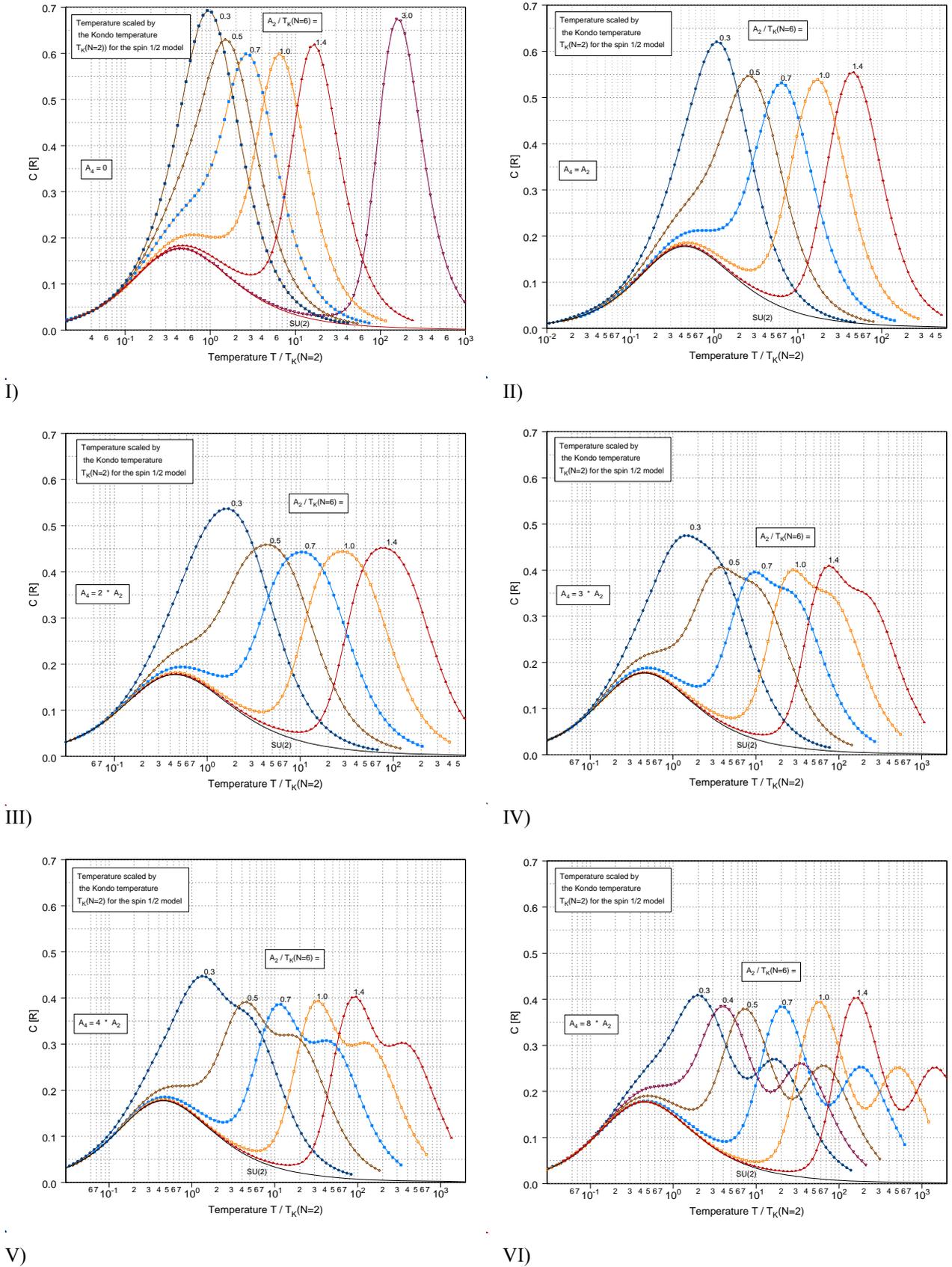

I)
II)
III)
IV)
V)
VI)

Fig. 5. (Color online) Specific heat as a function of T scaled by the respective Kondo temperature $T_K(N=2)$ of the effective spin-1/2 system at low temperatures: $C = \pi/3\ T/T_K(N=2)$. $A_4/A_2$ is kept fixed for each graph.